\documentclass[12pt]{article}

\newcommand{\beq}{\begin{equation}}
\newcommand{\eeq}{\end{equation}}

\newcommand{\Dslash}{/\!\!\!\!D}
\newcommand{\vslash}{/\!\!\!v}

\begin{document}
\vspace*{-.6in}
\thispagestyle{empty}
\begin{flushright}
CALT-68-2137\\
DOE RESEARCH AND\\
DEVELOPMENT REPORT
\end{flushright}
\baselineskip = 20pt
\vspace{.5in}
{\Large
\begin{center}
The b-quark and Symmetries of the Strong Interaction \footnote{Work supported
in part by the U.S. Dept. of Energy under Grant No. DE-FG03-92-ER40701.}
\end{center}}
\vspace{.4in}

\begin{center}
Mark B. Wise\\
\emph{California Institute of Technology, Pasadena, CA  91125 USA}
\end{center}
\begin{center}
\small{(Talk presented at the Twenty Beautiful Years of Bottom Physics
Symposium, Chicago, IL, 1997.)}
\end{center}
\vspace{1in}

\begin{center}
\textbf{abstract}
\end{center}
\begin{quotation}
\noindent
Applications of HQET and NRQCD to fragmentation are briefly reviewed.  The
special role of the b-quark in applications of heavy quark symmetry is
discussed.  Predictions of HQET for semileptonic B decays to excited charmed
mesons are considered.
\end{quotation}
\vfil

\newpage

\section{HQET}

\pagenumbering{arabic}

The heavy quark effective theory (HQET) is a limit of the theory of the strong
interactions appropriate for hadrons containing a single heavy quark $Q$.  In
such hadrons the light degrees of freedom typically have momentum of order
$\Lambda_{QCD}$.  Interactions of the heavy quark with the light degrees of
freedom cause changes in its four-velocity $v$ of order $\Delta v \sim
\Lambda_{QCD}/m_Q$.  Consequently for these hadrons it is a reasonable
approximation to take the limit of QCD where $m_Q \rightarrow \infty$ with the
heavy quark's four-velocity fixed.

The part of the QCD Lagrange density involving the heavy quark field is
\begin{equation}\label{1}
{\cal L} = \bar Q (i\Dslash - m_Q) Q.
\end{equation}
The QCD heavy quark field is related to its HQET counterpart by
\begin{equation}\label{2}
Q = e^{-im_{Q} v \cdot x} \left[1 + {i\Dslash\over 2m_Q} + \ldots\right] Q_v,
\end{equation}
where
\begin{equation}\label{3}
\vslash Q_v = Q_v.
\end{equation}
Putting Eq.~(\ref{2}) into the QCD Lagrange density and using eq.~(\ref{3})
yields
\begin{equation}\label{4}
{\cal L} = {\cal L}_{HQET} + \delta_1 {\cal L} + \ldots,
\end{equation}
where the HQET Lagrange density is~\cite{eichten1}
\begin{equation}\label{5}
{\cal L}_{HQET} = \bar Q_v i v \cdot D Q_v.
\end{equation}
If there are several heavy flavors a sum over different flavors of heavy quarks
is understood.  This Lagrange density is independent of the heavy quark mass
and spin and has the spin-flavor symmetry~\cite{isgur1} of HQET.  $\delta_1
{\cal L}$ contains corrections to the $m_Q \rightarrow \infty$ limit suppressed
by a single power of the heavy quark mass.  Explicitly~\cite{eichten2}
\begin{equation}\label{6}
\delta_1 {\cal L} = {1\over 2m_Q} [O_{kin,v}^{(Q)} + O_{mag,v}^{(Q)}],
\end{equation}
where the kinetic energy term is
\begin{equation}\label{7}
O_{kin,v}^{(Q)} = \bar Q_v (i D_\perp)^2 Q_v.
\end{equation}
Here, $D_\perp^\mu = D^\mu - v^\mu (v \cdot D)$ are the components of the
covariant derivative perpendicular to the four-velocity.  The chromomagnetic
energy term is
\begin{equation}\label{8}
O_{mag,v}^{(Q)} = \bar Q_v {g\over 2} \sigma_{\alpha\beta} G^{\alpha\beta A}
T^A Q_v.
\end{equation}
Note that the part of $\delta_1{\cal L}$ involving $O_{kin,v}^{(Q)}$ breaks the
flavor symmetry but not the spin symmetry.  $O_{mag,v}^{(Q)}$ breaks both
symmetries.

In the limit $m_Q \rightarrow \infty$ the angular momentum
 of the light degrees of freedom,
\begin{equation}\label{9}
\vec S_\ell = \vec J - \vec S_Q,
\end{equation}
is conserved~\cite{isgur2}.  Therefore, in this limit, hadrons occur in
doublets with total angular momentum
\[
j_\pm = s_\ell \pm 1/2.
\]
Here $\vec J^2 = j (j + 1)$ and $\vec S_\ell^2 = s_\ell (s_\ell + 1)$.  In the
case of mesons with $Q\bar q$ flavor quantum numbers, the ground state doublet
has spin-parity of the light degrees of freedom $s_\ell^{\pi_{\ell}} = {1\over
2}^-$.  For $Q = c$ this doublet contains the $D$ and $D^*$ mesons with spin 0
and 1 respectively and for $Q = b$ they are the $B$ and $B^*$ mesons.  An
excited doublet of mesons with $s_\ell^{\pi_{\ell}} = {3\over 2}^+$ has also
been observed.  In the $Q = c$ case this doublet contains the $D_1 (2420)$ and
$D_2^* (2460)$ with spin 1 and spin 2 respectively.  The analogous $Q = b$
mesons are called $B_1$ and $B_2^*$.

\section{NRQCD}

For quarkonia (i.e., $Q\bar Q$ hadrons) physical properties are usually
predicted using an expansion in $v/c$ where $v$ is the magnitude of the heavy
quarks' relative velocity and $c$ is the speed of light~\cite{bodwin}.  So the
appropriate limit of QCD to take in this case is the $c \rightarrow \infty$
limit~\cite{grinstein}.  In eq.~(\ref{1}) the speed of light was set to unity.
Making the factors of $c$ explicit it becomes
\begin{equation}\label{10}
{\cal L} = c \bar Q (i\Dslash - m_Q c) Q,
\end{equation}
where
\begin{equation}\label{11}
\partial_0 = {1\over c} {\partial\over\partial t},
\end{equation}
and the covariant derivative
\begin{equation}\label{12}
D_\mu = \partial_\mu + {ig\over c} A_\mu^A T^A.
\end{equation}
Note that the strong coupling $g$ has the same units as $\sqrt{c}$.  The full
QCD heavy quark field $Q$ is related to its NRQCD counterpart by
\begin{equation}\label{13}
Q = e^{-im_{Q} c^{2} t} \left[1 + {i\Dslash_\perp\over 2m_Q c} + \ldots
\right]\left( {\psi\atop 0} \right),
\end{equation}
where $\psi$ is a two component Pauli spinor and $D_\perp = (0, {\bf
D}_\perp)$.  Putting eq.~(\ref{13}) into eq.~(\ref{10}) gives
\begin{equation}\label{14}
{\cal L} = {\cal L}_{NRQCD} + \ldots,
\end{equation}
where
\begin{equation}\label{15}
{\cal L}_{NRQCD} = \psi^\dagger \left(i \left({\partial\over\partial t} + ig
A_0^A T^A\right) + {\vec\nabla^2\over 2m_Q} \right) \psi.
\end{equation}
The $c \rightarrow \infty$ limit of QCD is called non-relativistic quantum
chromodynamics (NRQCD).  Since the kinetic energy appears as a leading term in
NRQCD this theory does not have a heavy quark flavor symmetry; however, it
still has a heavy quark spin symmetry.  The gluon field $A_0$ in eq.~(\ref{15})
is not a propagating field.  It gives rise to a Coulomb potential between the
heavy quarks.  All the interactions of the propagating transverse gluons with
the heavy quarks are suppressed by powers of $1/c$.  The leading interaction of
the propagating transverse gluons with the heavy quarks is also invariant under
heavy quark spin symmetry.

\section{Special Role of the Bottom Quark}

The $c, b$ and $t$ quarks can be considered heavy.  Unfortunately the top is so
heavy that it decays before forming a hadron.  Heavy quark symmetry is not a
useful concept for the $t$-quark.  The charm quark mass is not large enough for
one to be confident that predictions based on heavy quark symmetry will work
well.  For charmonium $v^2/c^2 \sim 1/3$ and $\Lambda_{QCD}/m_c \sim 1/7$.
However, for the b-quark, corrections to predictions based on heavy quark
symmetry should be small.  This ``special role'' of the b-quark is illustrated
nicely by comparing with experiment the predictions of heavy quark symmetry for
fragmentation.

Heavy quark symmetry implies that the probability $P_{h_{Q} \rightarrow
h_{s}}^{(H)}$ for heavy quark $Q$ with spin along the fragmentation axis
(i.e., helicity) $h_Q$ to fragment to a hadron $H$ with spin of the light
degrees $s_\ell$, total spin $s$
 and helicity $h_s$ is~\cite{falk}
\begin{equation}\label{16}
P_{h_{Q} \rightarrow h_{s}}^{(H)} = P_{Q \rightarrow s_{\ell}} p_{h_{\ell}}
|\langle s_Q, h_Q; s_\ell, h_\ell| s, h_s \rangle |^2.
\end{equation}
In eq.~(\ref{16}) $P_{Q \rightarrow s_{\ell}}$ is the probability for the heavy
quark to fragment into the doublet with spin of the light degrees of freedom
$s_\ell$.  $p_{h_{\ell}}$ is the probability for the helicity of the light
degrees of freedom to be $h_\ell = h_s - h_Q$, given that the heavy quark
fragments to this doublet.  Parity invariance of the strong interactions
implies that
\begin{equation}\label{17}
p_{h_{\ell}} = p_{-h_{\ell}},
\end{equation}
and the definition of a probability implies that
\begin{equation}\label{18}
\sum_{h_{\ell}} p_{h_{\ell}} = 1.
\end{equation}
The constraints in eqs.~(\ref{18}) and~(\ref{17}) imply that there are
$s_\ell-1/2$ independent probabilities $p_{h_{\ell}}$.

For the $c\bar q$ ground state meson doublet $p_{1/2} = p_{-1/2} = 1/2$ and the
relative fragmentation probabilities are
\begin{equation}\label{19}
\begin{array}{ccccccc}
P_{1/2 \rightarrow 0}^{(D)} &:& P_{1/2 \rightarrow 1}^{(D^{*})} &:& P_{1/2
 \rightarrow 0}^{(D^{*})} &:& P_{1/2 \rightarrow -1}^{(D^{*})} \\[10pt]
{1\over 4} &:& {1 \over 2} &:& {1\over 4} &:& 0
\end{array}
\end{equation}
%\[
%P_{1/2 \rightarrow 0}^{(D)} \quad : \quad P_{1/2 \rightarrow 1}^{(D^{*})} 
% \quad : \quad P_{1/2 \rightarrow 0}^{(D^{*})} \quad : \quad P_{1/2 
% \rightarrow -1}^{(D^{*})} \]
%\begin{equation}\label{19}
%\qquad {1\over 4} \qquad :\qquad~   {1 \over 2} \qquad :\qquad {1\over 4} %\qquad:\qquad
%0.
%\end{equation}
For the excited $s_\ell^{\pi_{\ell}} = {3 \over 2}^+$ doublet the relative
fragmentation probabilities can be expressed using eq.~(\ref{16}) in terms of
$w_{3/2}$.  This parameter is defined by $p_{3/2} = p_{- 3/2} = (1/2)
{}~w_{3/2}$ and $p_{1/2} = p_{- 1/2} = (1/2) (1 - w_{3/2})$. 

In the charm system only part of eq.~(\ref{19}) is in agreement with
experiment. While the experimental
value for the relative probability to fragment to longitudinal and
transverse $D^*$ helicities agrees with
eq.~(\ref{19}), the experimental values for the 
probabilities to fragment to $D$ and $D^*$ are approximately equal~\cite{falk}
instead of in the ratio 1:3 that eq.~(\ref{19}) predicts. This discrepancy
is probably due to the $D^*$-$D$ mass difference which suppresses
fragmentation to the $D^*$. Recent LEP data
shows that predictions for fragmentation based on heavy quark symmetry work
better in the b-quark case~\cite{eigen}.  The experimental value for the probabilities to fragment
to the $B$ and $B^*$ are in the ratio 1:3 .  

Experimental information on $D^{**}$ production provides
the bound, $w_{3/2} < 0.24$~\cite{falk}. It would be very interesting to have
an experimental determination of the Falk-Peskin fragmentation parameter
 $w_{3/2}$.

Heavy quark spin symmetry also makes predictions for the alignment of quarkonia
produced by gluon fragmentation.  At leading order $v/c$ the gluon fragments to
$Q\bar Q$ in a color singlet configuration.  Two hard gluons occur in the final
state to conserve color and charge conjugation
, giving a fragmentation probability to
$^3S_1$ quarkonia of order $(\alpha_s (m_Q)/\pi)^3 (v/c)^3$.  However, a term
higher order in $v/c$ is much more important because it is lower order in
$\alpha_s (m_Q)/\pi$.  The gluon can fragment to the $Q\bar Q$ pair in a color
octet with two soft propagating NRQCD gluons in the final state (each with
typical momentum of order $m_Q v (v/c)$ in the quarkonium rest frame).  This
color octet process~\cite{braaten} gives a contribution to the $^3S_1$
fragmentation probability of order $(\alpha_s (m_Q)/\pi) (v/c)^7$.  The
fragmenting gluon has large energy (compared with $m_Q$) and is almost real.
Real gluons are transversely aligned.  Because the leading interactions of the
NRQCD propagating gluons preserve spin symmetry the final state $^3 S_1$
quarkonium is also transversely aligned~\cite{cho}.  (There are $\alpha_s (m_Q)$
and $v/c$ corrections~\cite{beneke1} that reduce this alignment.)  It may be
possible to test this prediction in the $Q = c$ case from large $p_\perp$ data
on $J/\psi$ and $\psi'$ production at the Tevatron~\cite{beneke2}.

\section{$B \rightarrow D_1 (2420) \lowercase{e}\bar\nu_{\lowercase{e}}$ and $B
\rightarrow D_2^* (2460) \lowercase{e}\bar\nu_{\lowercase{e}}$ Decay}

Semileptonic B decays have been extensively studied.  The semileptonic decays
$B \rightarrow D e\bar\nu_e$ and $B \rightarrow D^* e\bar\nu_e$ have branching
ratios of $(1.8 \pm 0.4)\%$ and $(4.6 \pm 0.3)\%$ respectively~\cite{particle}.
 They amount to about 60\% of the semileptonic decays.  The differential decay
rates are determined by matrix elements of the $b \rightarrow c$ weak
axial-vector and vector currents.  These matrix elements are usually written in
terms of Lorentz scalar form factors and the differential decay rates are
expressed in terms of them.  For comparisons with the predictions of HQET it is
convenient to write the form factors in terms of $w = v \cdot v'$.  In the
limit $m_Q \rightarrow \infty$ heavy quark spin symmetry implies that all six
form factors can be written in terms of a single function of $w$~\cite{isgur1}.
 Furthermore, heavy quark flavor symmetry implies that this function is
normalized to unity~\cite{isgur1,nussinov} at zero recoil, $w = 1$.  The
success of these predictions~\cite{cleo} indicates that in this case treating
the charm quark mass as large is a reasonable approximation.  At order
$1/m_{c,b}$ several new functions occur but the normalization of the zero
recoil matrix elements is preserved.

In the $m_Q \rightarrow \infty$ limit zero recoil matrix elements of the weak
axial vector and vector currents from the B-meson to any excited charmed meson
vanish because of heavy quark spin symmetry.  Since most of the phase space for
such decays is near zero recoil (e.g., for B decay to the $s_\ell^{\pi_{\ell}}
= {3\over 2}^+$ mesons $D_1 (2420)$ and $D_2^*(2460), 1< w < 1.3)$ the
$\Lambda_{QCD}/m_{c,b}$ corrections are very important.

The decay $B \rightarrow D_1 e\bar\nu_e$ has been observed.  CLEO and ALEPH,
respectively, find the branching ratios~\cite{aleph} $Br (B \rightarrow D_1
e\bar\nu_e) = (0.49 \pm 0.14)\%$ and $(0.74 \pm 0.16)\%$.  For $Br
(B\rightarrow D_2^* e\bar\nu_e)$ there are only upper limits.

The form factors that parametrize the $B \rightarrow D_1$ and $B \rightarrow
D_2^*$ matrix elements of the weak currents $V^\mu = \bar c \gamma^\mu b$ and
$A^\mu = \bar c \gamma^\mu \gamma_5 b$ are defined by
\begin{eqnarray}\label{21}
{\langle D_1 (v',\varepsilon)| V^\mu| B(v)\rangle\over \sqrt{m_{D_{1}} m_B}}
&=& f_{V_{1}} \varepsilon^{*\mu} + (f_{V_{2}} v^\mu + f_{V_{3}} v^{\prime\mu})
(\varepsilon^* \cdot v),\nonumber \\
{\langle D_1 (v',\varepsilon)| A^\mu| B(v)\rangle\over \sqrt{m_{D_{1}} m_B}}
&=& if_A \varepsilon^{\mu\alpha\beta\gamma} \varepsilon_\alpha^* v_\beta
v'\gamma,\nonumber \\
{\langle D_2^* (v',\varepsilon)| A^\mu| B(v)\rangle\over \sqrt{m_{D_{2}^{*}}
m_B}} &=& k_{A_{1}} \varepsilon^{*\mu\alpha}  v_\alpha + (k_{A_{2}} v^\mu +
k_{A_{3}} v^{\prime\mu})  \varepsilon^*_{\alpha\beta} v^\alpha
v^\beta,\nonumber \\
{\langle D_2^* (v',\varepsilon)| V^\mu| B(v)\rangle\over \sqrt{m_{D_{2}^{*}}
m_B}} &=&  ik_V \varepsilon^{\mu\alpha\beta\gamma} \varepsilon^*_{\alpha\sigma}
v^\sigma v_\beta v'_\gamma.
\end{eqnarray}
The form factors  $f_i$ and $k_i$ are functions of $w$.  In the $m_{c,b}
\rightarrow \infty$ limit they can be written in terms of a single function
$\tau (w)$~\cite{isgur3},
\begin{eqnarray}\label{22}
\sqrt{6} f_A &=& - (w + 1)\tau, \quad k_V = - \tau,\nonumber \\
\sqrt{6} f_{V_{1}} &=&  (1 - w^2)\tau, \quad k_{A_{1}} = - (1 +
w)\tau,\nonumber \\
\sqrt{6} f_{V_{2}} &=& - 3\tau, \quad k_{A_{2}} =  0,\nonumber \\
\sqrt{6} f_{V_{3}} &=&  (w - 2)\tau, \quad k_{A_{3}} =  \tau.\nonumber \\
\end{eqnarray}
Only the form factor $f_{V_{1}}$ contributes at zero recoil.  Surprisingly one
can predict its value~\cite{leibovich}
\begin{equation}\label{23}
\sqrt{6} f_{V_{1}} (1) = - {4 (\bar\Lambda' - \bar\Lambda)\tau (1)\over m_c},
\end{equation}
in terms of the $m_{c,b} \rightarrow \infty$ Isgur--Wise function $\tau$ and
the difference between the mass of the light degrees of freedom in the excited
$s_\ell^{\pi_{\ell}} = {3 \over 2}^+$ doublet $\bar\Lambda'$ and the mass of the light
degrees of freedom in the ground state doublet $\bar\Lambda$.  Experimentally
the difference $\bar\Lambda' - \bar\Lambda \simeq 0.39$ GeV.  (It can be
expressed in terms of measured hadron masses.)  A detailed discussion of the
$1/m_{c,b}$ corrections to these decays can be found in Refs. [18].  They
enhance the rate to $B \rightarrow D_1 e\bar\nu_e$ (compared with the $m_{c,b}
\rightarrow \infty$ limit) and lead to the expectation that its branching ratio
is greater than that for $B \rightarrow D_2^* e\bar\nu_e$.  This may explain
why semileptonic decays to the $D_2^*$ have not been observed.

\end{document}